\documentclass[aps,prb,article,twocolumn,showpacs,preprintnumbers,amsmath,amssymb,superscriptaddress]{revtex4-1}
\date{\today}
\usepackage{epsfig}
\usepackage{subfigure}
\usepackage{graphicx}
\usepackage{dcolumn}
\usepackage{bm}
\usepackage[colorlinks,linkcolor=blue,hyperindex,CJKbookmarks]{hyperref}
\usepackage{float}
\usepackage{hyperref}
\usepackage{comment}
\usepackage{color}

\begin{document}

\title{Photoemission Spectrum of Ca$_2$RuO$_4$:\\ Spin Polaron Physics in an $S=1$ Antiferromagnet with Anisotropies}

\author{Adam K\l{}'osi\'n"ski}
\affiliation{Institute of Theoretical Physics, Faculty of Physics, University of Warsaw, Pasteura 5, PL-02093 Warsaw, Poland}
\email[]{adam.klosinski@fuw.edu.pl}

\author{Dmitri V. Efremov}
 \affiliation{IFW Dresden, Helmholtzstr. 20, 01069 Dresden, Germany}

\author{Jeroen van den Brink}
 \affiliation{IFW Dresden, Helmholtzstr. 20, 01069 Dresden, Germany}
  \affiliation{Department of Physics, Technical University Dresden, Helmholtzstr. 10, 01069 Dresden, Germany}

\author{Krzysztof Wohlfeld}
\affiliation{Institute of Theoretical Physics, Faculty of Physics, University of Warsaw, Pasteura 5, PL-02093 Warsaw, Poland}

\date{\today}
\begin{abstract}
We derive an $S=1$ spin polaron model which describes the motion of a single hole introduced into the $S=1$ spin antiferromagnetic ground state
of Ca$_2$RuO$_4$. We solve the model using the self-consistent Born approximation and show that its hole spectral function qualitatively agrees with
the experimentally observed high-binding energy part of the Ca$_2$RuO$_4$ photoemission spectrum. We explain the observed 
peculiarities of the photoemission spectrum by linking them to two anisotropies present in the employed model---the spin anisotropy and the hopping anisotropy.
We verify that these anisotropies, and \emph{not} the possible differences between the ruthenate ($S=1$) and the cuprate ($S=1/2$) spin polaron models, 
are responsible for the strong qualitative differences between the photoemission spectrum of Ca$_2$RuO$_4$ and of the undoped cuprates.
\end{abstract}
\pacs{xxxx}
\maketitle

\section{Introduction.} \label{sec:introduction}

Understanding the strongly correlated physics of the transition metal oxides constitutes a nontrivial task~\cite{Dagotto1994, Imada1998, Lee2006, Khomskii2010, Khomskii2014}. 
On one hand, this is due to the fact that even the simplest, but still realistic, effective models may have to contain several degrees of freedom. On the other, this is due to the fact that such relatively simple models are often not solvable in the thermodynamic limit. That is why examples when a theoretical model can be solved without too drastic
approximations {\it and} explain the experimentally observed features of a correlated oxide are of interest.
One such case, known already since the end of the 80s of the last century, is the so-called spin polaron 
problem~\cite{Schmitt1988, Martinez1991, Ramsak1998, Manousakis2007, Wang2015, Grusdt2018, Bieniasz2019},
which explains the peak dispersion found in the photoemission spectrum of the (spin $S=1/2$) antiferromagnetically ordered and Mott insulating copper oxides~\cite{Wells1995, LaRosa1997, Kim1998, Damascelli2003, Shen2007}:
It turns out that the peak dispersion found in the spectra of the `parent compounds' to the high-temperature superconductors can be well explained using a $t$--$J$ or Hubbard model that is mapped onto a (spin) polaron problem. 

Interestingly, despite the still unresolved mysteries associated with high-temperature superconductivity, the copper oxides
are the simplest class of oxides to model. This is basically due to the fact that
the uncorrelated part of their physics can be effectively described 
using a single-band picture~\cite{Zhang1988, Lee2006}. Such situation is not realised in many other oxides, such as the manganites, vanadates, nickelates---or the 
ruthenates studied here~\cite{Imada1998, Khomskii2014}. In this case the effective models are far more involved and are never of the single-band $t$--$J$ or Hubbard variety. Can one thus expect that some of the principles of the spin polaron physics do hold there?

To investigate this rather general problem, we take a closer look at one of the intensively investigated transition metal oxides---Ca$_2$RuO$_4$~\cite{Alexander1999, Nakatsuji2000, Mizokawa2001, 
Lee2002, Gorelov2010, Kunkemoeller2015, Fatuzzo2015, Kunkemoeller2017, Jain2017, Sutter2017, Zhang2017, Das2018, Ricco2018, Pincini2019, Gretarsson2019} which
is a spin $S=1$ antiferromagnetically ordered~\cite{Mizokawa2001, Kunkemoeller2015, Kunkemoeller2017, Das2018, Pincini2019}  and Mott insulating analogue of the unconventional superconductor Sr$_2$RuO$_4$~\cite{Maeno1994}. 
The other reason for choosing this system is that recently its detailed photoemission spectrum was not only studied experimentally 
but also successfully modelled using a multiband Hubbard model~\cite{Sutter2017}. Nevertheless, as the multiband 
Hubbard model was solved using the single-site dynamical mean-field theory approach, it is not clear to what extent the spin polaron physics is present there. 

\begin{figure}[t!]
\includegraphics[keepaspectratio=true,width=0.5\textwidth]{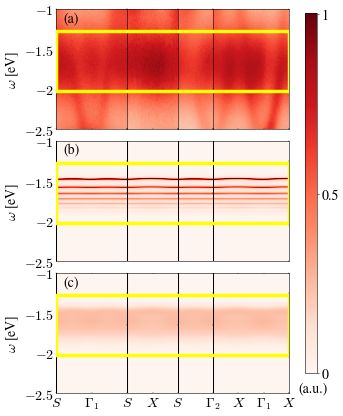}
  \caption{Comparison between the experimental and theoretical spectral functions of Ca$_2$RuO$_4$:
    (a) Angle resolved photoemission (ARPES) spectrum obtained for Ca$_2$RuO$_4$ and published in Ref.~\onlinecite{Sutter2017};
  (b) Hole spectral function $A({\bf k}, \omega)$ calculated for the spin $S=1$ $t$--$J$ Hamiltonian (\ref{model1}-\ref{model2})
   with ${\bf e}_{xz} = \hat{x}$, ${\bf e}_{yz}=\hat{y}$ and using the mapping onto the spin-polaronic model (\ref{finalhamiltonian}) and the SCBA method (see text);
  model (\ref{finalhamiltonian}) parameters: $t = 22 J$, $\epsilon = 5.6 J$, $\gamma = 0.25 J$, $J = 5.6 $ meV,
  numerical broadening of $A({\bf k}, \omega)$ $\delta = 1.1 J $;
  (c) Hole spectral function $A({\bf k}, \omega)$ calculated as in (b) but convoluted with a Gaussian with the half-width at half maximum equal to $0.5t$, 
  simulating the experimental resolution of ARPES on Ca$_2$RuO$_4$~\cite{Sutter2017}.
  The yellow rectangles mark the high binding energy parts of the spectra
  that are incoherent and almost momentum-independent, 
  are identified in ARPES as having a dominant $xz$/$yz$ orbital character~\cite{Sutter2017}, 
  and are theoretically modelled by panel (b) [The dispersive branch visible in (a), both inside and outside of the yellow rectangle and not discussed here, is associated with the $xy$ orbital \cite{Sutter2017}. See main text of the paper].
  Theoretical spectra (b-c) are normalised in the same manner as the ARPES spectrum [(a)] of Ref.~\onlinecite{Sutter2017}.
}
  \label{result}
\end{figure}

In this paper we concentrate on the origin of the incoherent 
and almost momentum-independent Ca$_2$RuO$_4$ photoemission
spectrum, that is present at the high-binding energy part [see the yellow rectangle of Fig.~\ref{result}(a)] and is associated with the hole motion in the $xz$ and $yz$ orbitals~\cite{Sutter2017}.
The reason for leaving the $xy$ orbital out of our analysis is that the highly dispersive, quasiparticle-like part of the photoemission spectrum stretching from low- to high-binding energy [see Fig.~\ref{result}(a)] can be easily understood as the free hole motion on the $xy$ orbitals. This follows straightforwardly from, on the one hand, the relatively large energy gap between the $xy$ and $xz/yz$ orbitals leading to the $xy$ orbital being fully occupied and, on the other hand, the absence of any mixing between $xy$ and $xz/yz$ orbitals, be it from hopping or the Coulomb interaction. Thus, including the $xy$ orbitals in our analysis would not much enrich our understanding of the physics at work in this compound.

Altogether, our aim here is twofold:
First, we want to model the high-binding energy part of the photoemission spectrum of Ca$_2$RuO$_4$ by using a realistic {\it $S=1$ spin polaron} model, specifically derived for this case.
Second, we wish to understand to what extent such an $S=1$ spin polaron problem is different from the `standard' (i.e. $S=1/2$) spin polaron model that is well-known from the cuprate studies.

The paper is organized as follows. In Sec.~\ref{sec:tjmodel} we introduce the $t$-$J$ model that describes the motion of a single hole in the ground state of Ca$_2$RuO$_4$.  Next, in Sec.~\ref{sec:fromto} we perform a mapping of the $t$-$J$ model onto a S=1 spin polaron model. The latter is solved using the self-consistent Born approximation (SCBA) in Sec.~\ref{sec:mandr}. Finally, we discuss the obtained results in Sec.~\ref{sec:discussion} and end the paper with conclusions in Sec.~\ref{sec:conclusions}. The paper is supplemented by an Appendix which contains some details of the mapping from the $t$-$J$ to the polaron model.

\section{${\bf t}$-${\bf J}$ model.} \label{sec:tjmodel}

In order to model (the high-binding energy part of) the photoemission spectrum of Ca$_2$RuO$_4$ we follow the scheme that was already developed ca. 30 years ago
and, as mentioned in the introduction, was successfully used to describe, {\it inter alia}, the photoemission spectra of several undoped
copper oxides~\cite{Wells1995, LaRosa1997, Kim1998, Damascelli2003, Shen2007}. Thus, we consider 
an appropriate $t$--$J$-like Hamiltonian constructed as a sum of two parts ${\mathcal H} = {\mathcal H}_J+ {\mathcal H}_t$. 

The first part, ${\mathcal H}_J$, describes the low energy physics of the Mott insulating Ca$_2$RuO$_4$ in terms of the interaction between the localised $S=1$ magnetic moments. The relevant Hamiltonian is well-known in this case and reads~\cite{Kunkemoeller2015, Kunkemoeller2017, Jain2017},
\begin{equation} \label{model1}
  {\mathcal H}_J = J \sum\limits_{\langle {\bf i}, {\bf j} \rangle} \mathbf{S_{\bf i}} \cdot \mathbf{S_{\bf j}} + \epsilon \sum\limits_{\bf i} \left( S_{\bf i}^z \right)^2 + \gamma \sum\limits_{\bf i} \left( S_{\bf i}^x \right)^2,
\end{equation}
where the summation runs over all nearest-neighbour pairs on a 2D square lattice, $J$ is the spin exchange constant, and $\mathbf{S_{\bf i}}$ are the spin $S=1$ operators. As already discussed in Refs.~\onlinecite{Kunkemoeller2015, Kunkemoeller2017, Jain2017} the spin model is highly anisotropic, with the suggested values of the $\hat{z}$ ($\hat{x}$) axis anisotropy being equal to 
$\epsilon = 5.6 J$ ($\gamma = 0.25 J$), respectively, with $J =  5.6$ meV reproducing the spin wave dispersion observed in the inelastic neutron scattering experiment~\cite{Kunkemoeller2015, Kunkemoeller2017, Jain2017}.
We note that such a large spin anisotropy originates in the large spin-orbit coupling on the ruthenium ions 
which, however, is quite widely considered as not strong enough to stabilise the $S=0$ ground state~\cite{Mizokawa2001, Kunkemoeller2015, Kunkemoeller2017, Das2018, Pincini2019}. Although the latter result can naively be understood as a consequence of the crystal field splitting (between the $xz, yz$ and the $xy$ orbitals) being about twice larger than the spin-orbit coupling~\cite{Das2018} and therefore the spin $S=0$ states having considerably higher energy than the spin $S=1$ states, see Fig.~S1 of~\cite{Jain2017}, it has been postulated~\cite{Khaliullin2013, Akbari2014, Jain2017, Gretarsson2019} that nevertheless the `excitonic magnetism' can be at play here.

The second part of the Hamiltonian, ${\mathcal H}_t$, is the `kinetic' term. This term is introduced, in order to describe the motion of a single hole created in the ruthenium oxide
plane in the photoemission experiment. We restrict our model to the $xz$ and $yz$ orbitals, for, as discussed in the introduction, we are solely interested in the part of the photoemission spectrum associated with these orbitals~\cite{Sutter2017}. Altogether, we end up with,
\begin{align} \label{model2}
  \begin{split}
    {\mathcal H}_t &= - t \sum\limits_{{\bf i},  \sigma} \left( \tilde{c}_{{\bf i}+{{{\bf e}_{xz}}},xz,\sigma}^{\dag} \tilde{c}_{{\bf i},xz,\sigma} +\tilde{c}_{{\bf i}+{{\bf e}_{yz}},yz,\sigma}^{\dag} \tilde{c}_{{\bf i},yz,\sigma} +h.c.\right),\\
  \end{split}
\end{align}
where the first (second) term describes the hoppings of an electron with spin $\sigma$ between the nearest neighbor ruthenium $xz$ ($yz$) orbitals along the ${\bf e}_{xz}=\hat{x}$ (${\bf e}_{yz}=\hat{y}$) direction in the 2D square lattice, respectively.
Such effectively one-dimensional (`directional') hoppings follow from the Slater-Koster scheme~\cite{Slater1954} applied to the square lattice geometry of the ruthenium oxide plane and is, in fact, a common feature of systems with active $\{xz, yz\}$ orbital degrees of freedom~\cite{Harris2004}. As for the value of the hopping element $t$ in Ca$_2$RuO$_4$ we choose $t= 123$ meV~\cite{Gorelov2010}, i.e. $t = 22J$ for the above chosen realistic value of $J=5.6$ meV. We note that to simplify the analysis we skip here the spin-orbit coupling between holes in the $xz$ and $yz$ orbitals. Such a simplification is not {\it a priori} justified for a realistic situation in Ca$_2$RuO$_4$ but is rationalised by 
the intuitive understanding of its spectral functions presented in Ref.~\onlinecite{Sutter2017}, which relies on the Hund's coupling and does not include the spin-orbit coupling as an essential part. Moreover, 
a (surprisingly) good agreement between the theoretical results presented below
and the experimental results, cf. Fig.~\ref{result}, {\it a posteriori} legitimizes this assumption further. Finally, it will not affect the study of the possible intrinsic differences between the $S=1/2$ and $S=1$ spin polaron models.

There are two projections in place in the kinetic 
Hamiltonian (\ref{model2}). First, due to the strong on-site Coulomb repulsion $U$--and since we confine ourselves to the low energy physics valid for energies smaller than the 
`Hubbard' $U$--we restrict the hole motion to the Hilbert space spanned by the $d^2$ or $d^1$ multiplets on the single ruthenium ions. [Since
the $xy$ orbital is considered to be `always' occupied by two electrons in the studied model~\cite{Mizokawa2001, Gorelov2010, Sutter2017}, the $xy$ electrons are integrated out and effectively the nominal occupancy 
of the ruthenium ions is not $d^4$ ($d^3$) but $d^2$ ($d^1$) in the undoped (single-hole) case, respectively.] 
As typical to any $t$--$J$-like model~\cite{Chao1977} to formally denote such a constraint we use the tildas above the electron creation and annihilation operators. Second, just as in the case of the ground state (see discussion above), we project the spin $S=0$ states out of the Hilbert space and, formally, Hamiltonian~(\ref{model2}) contains such projections. We will not, however, keep them explicit in the formulae below. 

Finally, as we are interested
in the photoemission spectrum, we define the following orbitally-resolved hole spectral function,
\begin{align}
A_{\alpha}({\bf k}, \omega)\!= \!- \frac{1}{\pi} 
{\rm Im} \Big\langle 0 \Big| \tilde{c}^\dag_{{\bf k}, \alpha, \sigma} \frac{1}{\omega - {\mathcal H} + E_0 +i \delta } \tilde{c}_{{\bf k}, \alpha, \sigma}  \Big| 0 \Big\rangle,
\end{align}
where $| 0 \rangle$ is the ground state of the undoped $t$--$J$ model (\ref{model1}-\ref{model2}) with energy $E_0$, 
$\delta$ is the infinitesimally small broadening that is nevertheless finite in the numerical calculations below,
and we explicitly keep the orbital index $\alpha \in \{ xz, yz \}$ but suppress the spin index $\sigma$ (the spectral function is spin-independent).
In what follows we are also interested in the orbitally-integrated spectral function which is defined in the usual way:
$A({\bf k}, \omega)=\sum_{\alpha}A_{\alpha}({\bf k}, \omega)$. 

\section{From ${\bf t}$-${\bf J}$ to polaron model.} \label{sec:fromto}

Stimulated by the successful description of the photoemission spectra of the undoped cuprates~\cite{Wells1995, LaRosa1997, Kim1998, Ramsak1998, Damascelli2003, Shen2007, Manousakis2007, Wang2015}
and to gain a better insight into the physics of the photoemission problem, we perform a mapping of the $S=1$ $t$--$J$ problem onto 
an $S=1$ spin polaron problem. This is done in two steps: 

First, we introduce the slave fermions,
\begin{equation} \label{slavefermions}
  \begin{array}{l}
    \tilde{c}_{{\bf i},\alpha,\uparrow} \rightarrow h_{{\bf i},\alpha}^{\dag}, \quad \quad
    \tilde{c}_{{\bf i},\alpha,\downarrow} \rightarrow \hat{A} \: h_{{\bf i},\alpha}^{\dag} \: S_{\bf i}^+,\\
  \end{array}
\end{equation}
where $h_{{\bf i},\alpha}^{\dag}$ is the creation operator for a spinless hole on site $i$ and orbital $\alpha$, $S_{\bf i}^+$ is the spin $S=1$ operator on site ${\bf i}$ and $\hat{A}$ is an operator yet to be determined. It can be shown that in the Hilbert space being considered, that is with the $S=0$ states projected out, the $\hat{A}$ operator is diagonal and an explicit expression for it can be found (see~Sec.~\ref{sec:appendix} for details). 
Second, we we rotate spins on 
one of the antiferromagnetic sublattices
and express the spin operators through bosonic operators by way of the Holstein-Primakoff transformation. Finally, we use the linear spin wave approximation and the Bogoliubov transformation to diagonalize the resulting spin Hamiltonian--see~Sec.~\ref{sec:appendix} for details. 

In the end we are left with a diagonal magnon term and a vertex coupling spinless holes to magnons in the following $S=1$ spin polaron Hamiltonian:
\begin{align} \label{finalhamiltonian}
    H &= H_t + H_J \approx \sum\limits_{\bf q} \: {\Omega}_{\bf q} \: {\beta}_{\bf q}^{\dag} {\beta}_{\bf q} + E_0 \nonumber \\
    &+\frac{\sqrt{2} \: t}{\sqrt{N}} \sum\limits_{\textbf{k},\textbf{q}}\Big[\left({\gamma}_{k_{x}}v_{\textbf{q}}+{\gamma}_{k_{x}-q_{x}}u_{\textbf{q}}\right)
    h_{\textbf{k},xz}^{\dag}h_{\textbf{k}-\textbf{q},xz} {\beta}_{\textbf{q}}\nonumber \\
    &+ \left({\gamma}_{k_{y}}v_{\textbf{q}}+{\gamma}_{k_{y}-q_{y}}u_{\textbf{q}}\right)
    h_{\textbf{k},yz}^{\dag}h_{\textbf{k}-\textbf{q},yz}{\beta}_{\textbf{q}}+h.c. \Big],
\end{align}
where $\gamma_{k_i} = \cos(k_i)$ and $\beta_q$ are the Bogoliubov boson (magnon) annihilation operators.
$u_{\bf q}, v_{\bf q}$ are the Bogoliubov coefficients - see Sec.~\ref{sec:appendix} for details.
The above transformations also lead directly to the expression for the spectral function in terms of the spinless hole Green's function (see~Sec.~\ref{sec:appendix} for details):
\begin{align}
A_{\alpha}({\bf k}, \omega)\! =\! - \frac{1}{\pi} 
{\rm Im} \Big\langle 0 \Big| {h}_{{\bf k}, \alpha} \frac{1}{\omega - H + E_0 +i \delta } {h}^\dag_{{\bf k}, \alpha}  \Big| 0 \Big\rangle.
\end{align}

\section{ Methods and results.} \label{sec:mandr}

We calculate the hole spectral function $A_{\alpha}({\bf k}, \omega)$ using the
self-consistent Born approximation (SCBA), see Ref.~\cite{Martinez1991}. Such approach has been widely-successful in 
obtaining the cuprate spectral functions~\cite{Martinez1991, Ramsak1998, Manousakis2007, Wang2015, Bieniasz2018} 
and amounts to neglecting the so-called crossing diagrams and summing
all the other (`rainbow diagrams') to infinite order. The resulting self-consistent expressions for the 
self-energies and the Green's function are given in~Sec.~\ref{sec:appendix}. These equations are then solved numerically on a finite
lattice of $36 \times 36$ ${\bf k}$ points. The resulting orbitally-integrated hole spectral function $A({\bf k}, \omega)$
is calculated for the realistic parameters of the model (see above) and is shown in Fig.~\ref{result}(b).

The calculated hole spectral function qualitatively reproduces the incoherent 
and almost momentum-independent spectrum observed in the high binding energy part of 
Ca$_2$RuO$_4$ photoemission results found in Ref.~\onlinecite{Sutter2017} and reproduced in Fig.~\ref{result}(a). Although the onset of several horizontal `stripes' in the theoretical spectrum (see below) make the 
similarities between the theoretical and experimental spectral functions less apparent at first sight [Fig.~\ref{result}(b)], a convolution of the theoretical spectral function with the available experimental
resolution of ca. $0.5t$ yields a spectrum [Fig.~\ref{result}(c)] which surprisingly well resembles the high binding energy part of the observed experimental spectrum [Fig.~\ref{result}(a)]: 
both spectra have an incoherent character, without a clear quasiparticle band emerging, and only very weak dependence of its intensity on the momentum. 
(A weak dependence on the momentum of the intensities in the high binding energy part of the observed experimental spectrum [Fig.~\ref{result}(a)] originates in the $xy$ orbital spectral function, not considered
here but explained in detail in Ref.~\onlinecite{Sutter2017}.)
What is more, the low- and high-energy edges of both broad spectra are basically momentum-independent. Finally, also the overall energy scale, which is given by the width of the broad spectrum estimated at the half maximum intensity, is of the same order of magnitude in both cases and amounts to about $0.5$ eV. 

\section{Discussion.} \label{sec:discussion}

What might be the origin of the onset of such an incoherent,
almost momentum-independent and, apart from the horizontal `stripes', rather featureless spectrum of Fig.~\ref{result}(b)? 
Looking first at model (\ref{model1}-\ref{model2}) we can immediately note what distinguishes it from the `standard' $S=1/2$ $t$--$J$ model, that
has been widely used to describe the photoemission spectra of the undoped cuprates~\cite{Schmitt1988, Martinez1991, Ramsak1998, Manousakis2007, Wang2015} and for which such a broad and flat incoherent band has not been observed. The most apparent are the two anisotropies. The spin anisotropy reflects the distortion of the lattice and leads to the $\gamma$ and $\epsilon$ terms in Eq.~(\ref{model1}). 
The perfect hopping anisotropy, on the other hand, 
which has its origin in the nominal valence of the ruthenium ions and the geometry of the ruthenium-oxide plane, leads to an effectively one-dimensional hole motion, cf. Eq.~(\ref{model2}). On top of that,
a more subtle distinction is related to the larger value of the spin $S=1$ in the studied model. The latter leads to the onset of additional projection operators in the hopping part of the Hamiltonian.

\begin{figure}[t!]
  \includegraphics[keepaspectratio=true,width=0.5\textwidth]{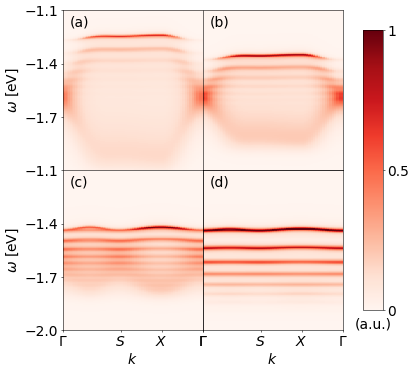}
  \caption{The hole spectral functions $A({\bf k}, \omega)$ obtained for distinct versions of the relevant $t$--$J$ models and calculated by mapping 
  onto  the spin-polaronic
  model and using the SCBA method (see text): (a) the `standard' spin $S=1/2$ $t$--$J$ model, cf. Ref.~\onlinecite{Martinez1991};
  (b) spin $S=1$ $t$--$J$ model with neither the spin nor the hopping anisotropy, i.e. model (\ref{model1}-\ref{model2}) with $\varepsilon=\gamma \equiv 0$ and ${\bf e}_{xz}\equiv{\bf e}_{yz} \in \{\hat{x}, \hat{y}\}$;
  (c) the spin $S=1$ $t$--$J$ model with the hopping anisotropy as suggested for Ca$_2$RuO$_4$  but no spin anisotropy, i.e. model (\ref{model1}-\ref{model2}) with ${\bf e}_{xz} = \hat{x}$ 
  and ${\bf e}_{yz}=\hat{y}$;
  and (d) the spin $S=1$ $t$--$J$ model with both anisotropies present as suggested for Ca$_2$RuO$_4$ and equivalent to Fig.~\ref{result}(b), 
  i.e. model (\ref{model1}-\ref{model2}) with ${\bf e}_{xz} = \hat{x}$, ${\bf e}_{yz}=\hat{y}$, and all model parameters as in Fig.~\ref{result}(b).
  All spectra normalised as Fig.~\ref{result}.
  }
  \label{fourplots}
\end{figure}
We explore the above-listed differences in detail by comparing the spectral function $A({\bf k}, \omega)$ calculated for the distinct versions of the relevant $t$--$J$ models, cf.~Fig. \ref{fourplots}. Firstly, it is evident that the effects of the spin anisotropy are profound, see Figs.~\ref{fourplots} (c) and (d).
On the one hand, it makes the spectrum less dispersive and in general more featureless; on the other it leads to the formation of a horizontal stripes superimposed onto the otherwise featureless spectrum. Both the former and the latter can be understood, when one considers the fact that the very large anisotropy limit leads, in this case, to the dominant Ising-like interactions between spins. 
This triggers the hole confinement in a linear string potential and leads to a well-known ladder-like spectrum with the horizontal `stripes'~\cite{Martinez1991, Dagotto1994, Bieniasz2018}.

\begin{figure}[t!]
  \includegraphics[keepaspectratio=true,width=0.5\textwidth]{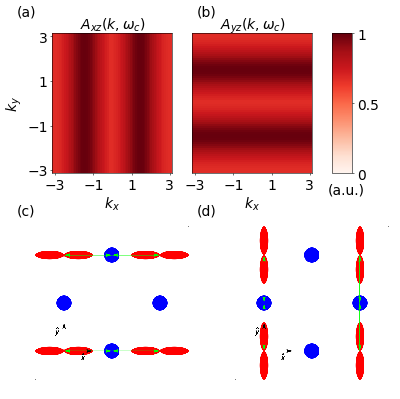}
  \caption{The one dimensional character of the orbitally-resolved hole spectral function:
  (a) Constant-energy cut of the spectral function $A_{xz} ({\bf k}, \omega_c)$ for a hole introduced into the $xz$ orbital ($\omega_c = -1.9 $ eV);
  (b) Constant-energy cut of the spectral function $A_{yz} ({\bf k}, \omega_c)$ for a hole introduced into the $yz$ orbital ($\omega_c = -1.9 $ eV);
  (c-d) A schematic view of the ruthenium-oxygen plane explaining the dominant one-dimensional character of the electronic
  hopping processes on the single-particle level that is also inherited by the many-body hopping processes of Eq.~(\ref{model2}): 
  For the $xz$ ($yz$) orbital, only hopping in the $\hat{x}$ ($\hat{y}$) direction is possible, cf. panel (c) [(d)]~\cite{Slater1954, Harris2004}.
  The oxygen (ruthenium) orbitals are shown in blue (red).}
  \label{oned}
\end{figure}
\begin{figure}[t!]
  \includegraphics[keepaspectratio=true,width=0.5\textwidth]{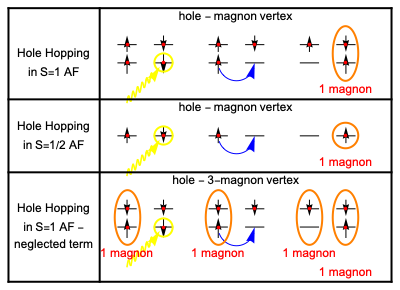}
  \caption{A schematic view of the possible nearest neighbor hole hoppings in the $S=1$ and $S=1/2$ antiferromagnet (AF): 
(Top panel) A hopping process in the $S=1$ antiferromagnet that, according to the here studied $t$--$J$ Hamiltonian 
(\ref{model1}-\ref{model2}), leads to the creation of one magnon in the effective $S=1$ spin polaron model~(\ref{finalhamiltonian});
(Middle panel) An analogous hopping process as above but in the $S=1/2$ antiferromagnet which, according to the `standard' $t$--$J$ 
Hamiltonian~\cite{Chao1977}, leads to the creation of one magnon in the spin polaron model of Ref.~\onlinecite{Martinez1991};
(Bottom panel) A hopping process in the $S=1$ antiferromagnet that, according to the here studied $t$--$J$ Hamiltonian 
(\ref{model1}-\ref{model2}), leads to the creation of three magnons and is {\it neglected} in the $S=1$ spin polaron model~(\ref{finalhamiltonian})
for it goes beyond the linear spin wave approximation.}
  \label{hoppingprocesses}
\end{figure}
Secondly, one can see that the one-dimensional hole motion completely changes the character of the spectral function, cf.~Fig.~\ref{fourplots}(b) and (c). 
In order to better understand why this happens, in Fig.~\ref{oned} we present the constant energy cuts of two spectral functions--one describing a hole in the $xz$ orbital, the other a hole in the $yz$ orbital. We see that the one-dimensional hole motion---a consequence of the geometry of the ruthenium-oxygen plane and the vanishing of the transfer integrals between the oxygen $p$ orbitals and some of the $t_{2g}$ orbitals~\cite{Harris2004}---is reflected in the hole spectral functions. They both show a manifestly one-dimensional dispersion, very much unlike what we see, for instance, in the copper oxides~\cite{Wells1995, LaRosa1997, Kim1998, Damascelli2003, Shen2007}.  We note that, while including a finite spin-orbit coupling for holes in the $xz$ or $yz$ orbitals would naturally lead to
the `mixing' between the one-dimensional bands, a good agreement between the theoretical and experimental spectra suggests that such an effect should be small in Ca$_2$RuO$_4$.

Finally, with all other parameters equal, the fact that we do not consider here a spin $S=1/2$ 
(which would be formed by a single hole or electron per site) but a spin $S=1$ antiferromagnet 
(two holes on each site) does not influence the spectral function qualitatively--thus, the difference between these two cases is purely quantitative, cf. Fig.~\ref{fourplots}(a) and (b).
To understand why it is so, let us compare the possible hole hopping processes in the $S=1/2$ and $S=1$ antiferromagnet, which are represented schematically in Fig.~\ref{hoppingprocesses}. What we can conclude by looking at the process represented on the bottom panel is that all the more complex processes, which have no analog in the single hole per site case, involve more than one magnon. In fact they involve either three or five magnons, which is why we exclude them in the spin wave approximation employed here 
and why they are absent from Hamiltonian~(\ref{finalhamiltonian}). Consequently, only the simplest process remains, the one analogous to the only process possible in the spin $S=1/2$ case, cf. the first two panels of Fig.~\ref{hoppingprocesses}. We stress that such a similarity between the hole moving in the $S=1/2$ and the $S=1$ antiferromagnet would not be achieved in the classical double exchange picture~\cite{Zener1951}, for the latter one would not allow for the existence of the $|1, 0 \rangle$ states on any site.

\section{Conclusions.} \label{sec:conclusions}

In this work we showed how a relatively simple spin $S=1$ $t$--$J$ model, that was mapped onto an $S=1$ spin polaron model, can qualitatively reproduce
 the high-binding energy part of the observed Ca$_2$RuO$_4$ photoemission spectrum. In particular, we were able to explain the observed 
incoherent and almost momentum-independent photoemission spectrum by linking these peculiar features of the spectrum to two anisotropies present 
in the employed spin polaron model---the spin anisotropy~\cite{Kunkemoeller2017} and the hopping anisotropy~\cite{Harris2004, Sutter2017}.

Interestingly, the differences between the spectral functions of the `standard' spin polaron model well-known from the cuprates (i.e. spin $S=1/2$) 
and the model for Ca$_2$RuO$_4$  
should not be regarded as being intrinsic to the $S=1$ spin polaron model: They are all solely related to the above-mentioned strong anisotropies
present in the model suggested for Ca$_2$RuO$_4$ and {\it not} to the potential differences between the hole moving in the $S=1/2$ and the $S=1$ antiferromagnet.
These turn out to be basically irrelevant in the linear spin wave approximation. Such a result can naturally be expected following basic quantum
mechanics but would not be achieved in the well-known double exchange picture~\cite{Zener1951}, for in that classical approach the hole would not be able to hop at all in 
the $S=1$ antiferromagnet. 

\section*{Acknowledgments.} 

We are very grateful to David Sutter and Johan Chang for sharing the experimental data that was presented in Ref.~\onlinecite{Sutter2017}
 and which allowed us to plot Fig.~\ref{result}(a). We thank Eugenio Paris for insightful discussions. A.K. thanks the IFW Dresden for the kind hospitality.
A.K. and K.W. (K.W.) acknowledge(s) support by Narodowe Centrum Nauki (NCN, Poland) under Projects No.~2016/22/E/ST3/00560 
(2016/23/B/ST3/00839), respectively. J.v.d.B. acknowledges financial support from the German Research Foundation (Deutsche Forschungsgemeinschaft, DFG) via 
SFB1143 project A5 and the W\"urzburg-Dresden Cluster of Excellence ct.qmat.

% APPENDIX
\section*{APPENDIX: Derivation of the Polaron model} \label{sec:appendix}

\subsection{Mapping onto a polaronic model: Spin Hamiltonian ($\mathcal{H}_J$)}

To diagonalize the spin $S=1$ Hamiltonian discussed in the paper [cf.~Eq.~(\ref{model1}) of the main text of the paper] we start by performing two rotations of spins. First, we make a different choice of the spin quantization axis $\hat{z}$ - in this case we pick the axis without an anisotropy term. Second, we perform a $\pi$ rotation of spins around the $\hat{x}$ axis on one of the two sublattices---such transformation maps the anticipated antiferromagnetic ground state onto a ferromagnetic state. The result is the rotated spin Hamiltonian
\begin{align} \label{eq:1}
   \mathcal{\tilde{H}}_J &= \sum\limits_{\left< {\bf i,j} \right>} J \: \left[ -\: S_{\bf i}^z S_{\bf j}^z + \frac{1}{2} \left( S_{\bf i}^+ S_{\bf j}^+ + S_{\bf i}^- S_j^- \right) \right] \nonumber \\
    &+ \gamma \sum\limits_{\bf i}\left(S_{\bf i}^y\right)^2 + \epsilon \sum\limits_{\bf i}\left(S_{\bf i}^x\right)^2.
\end{align}

The next step is to utilize the Holstein-Primakoff transformation and the linear spin wave approximation using the assumption that the ground state is ferromagnetic and dressed with magnons. The thus obtained Hamiltonian is then diagonalized using the successive Fourier and Bogoliubov transformations. The resulting spin Hamiltonian reads
\begin{align} \label{eq:1}
   {H}_J &= \sum_{\bf q}   {\Omega}_{\bf q} \beta^\dag_{\bf q} \beta_{\bf q} \; ,
\end{align}
where the magnon energies $ {\Omega}_{\bf q}$ are given by
\begin{equation}
    {\Omega}_{\bf q} = Y \sqrt{1 - \left( \frac{Z_{\bf q}}{Y} \right)^2},
\end{equation}
with
\begin{equation} \label{hj solved}
  \begin{array}{l}
    Y = 4J+{\epsilon}+{\gamma},\\
    Z_{\bf q} = 4J{\gamma}_{\bf q}+{\epsilon}-{\gamma},\\
  \end{array}
\end{equation}
with
\begin{equation}
    \gamma_{\bf q} = \frac{1}{2} \left( \cos(q_x) + \cos(q_y) \right).
\end{equation}
The Bogoliubov coefficients $u_{\bf q}$ and  $v_{\bf q}$, which describe the connection between the bosons before and after the Bogoliubov transformation and
are needed to correctly express the kinetic part of the Hamiltonian in the polaronic language (see next section), 
are expressed through the magnon energies  $ {\Omega}_{\bf q}$ via the standard formulae, cf. Ref.~\onlinecite{Martinez1991}.

\subsection{Mapping onto a polaronic model: Kinetic Hamiltonian ($\mathcal{H}_t$)}

\subsubsection{Dealing with projection operators}
The implicit projection operators in Eq.~(\ref{model2}) in the main text have a relatively complex form. The easiest way to deal with them is to construct the projected form of the two single orbital, two-site kinetic Hamiltonians
\begin{align} \label{hamt}
    (H_t^{xz})_{{\bf i},{\bf i}+{\bf e}_{xz}} &= -t \sum\limits_{\sigma} \tilde{c}_{{\bf i},xz,\sigma}^{\dag}\tilde{c}_{{\bf i}+{\bf e}_{xz},xz,\sigma} + h.c.,\\
    \label{hamt2}
    (H_t^{yz})_{{\bf i},{\bf i}+{\bf e}_{yz}} &= -t \sum\limits_{\sigma} \tilde{c}_{{\bf i},yz,\sigma}^{\dag}\tilde{c}_{{\bf i}+{\bf e}_{yz},yz,\sigma} + h.c.,
\end{align}
step by step, that is by considering every possible hopping process and the matrix element associated with it.

We consider a single hole introduced into the half-filled ground state and $H_t$ conserves the number of electrons in the system. 
Moreover, the `no double occupancy' constraint implies that there can never be 3 electrons on one site in the $xz/yz$ orbitals. 
As a result, a single hole introduced into the system propagates leaving the number of electrons in the $xz/yz$ orbitals on all other sites unchanged and equal to two. 
It follows that the only nonzero matrix elements of the two-site Hamiltonians (\ref{hamt}-\ref{hamt2}) describe processes in which a hole hops between a doubly occupied site and a singly occupied site. Furthermore, these hoppings obey two rules:
\begin{enumerate}
  \item The hole can hop to the neighbouring site, albeit only to an unoccupied orbital.
  \item The hole can hop to the neighbouring site, albeit only if the resulting on-site wavefunction is not a spin singlet.
\end{enumerate}
The first rule is simply the `no double occupancy' constraint and the second rule follows from the exclusion of the $S=0$ sector of the Hilbert space discussed in the main text. 

All of the above means that on each site we have either the spin triplet or one spin $S=1/2$ fermion on the $xz/yz$ orbitals. This in turn implies that for each pair of neighbouring sites ${\bf i,j}$ we can use the following basis
\begin{equation} \label{site basis}
\left\{ \Big| {^{\uparrow}_{\_}} \Big>_{\bf i},  \Big| {^{\downarrow}_{\_}} \Big>_{\bf i}, \Big| {^{\_}_{\uparrow}} \Big>_{\bf i}, \Big| {^{\_}_{\downarrow}} \Big>_{\bf i} \right\} \; \otimes \; \left\{ \Big| {^{\downarrow}_{\downarrow}} \Big>_{\bf j}, \Big| {^{\uparrow}_{\uparrow}} \Big>_{\bf j}, \frac{1}{\sqrt{2}} \left( \Big| {^{\uparrow}_{\downarrow}} \Big> + \Big| {^{\downarrow}_{\uparrow}} \Big> \right)_{ \bf j} \right\},
\end{equation}
where the upper and lower positions are the two ($xz/yz$) orbitals, the arrows show spin and the two particle states form the standard triplet. The main task now is to consider the matrix elements of (\ref{hamt}-\ref{hamt2}) in the basis (\ref{site basis}). 

Because the Hilbert space is 12-dimensional there are 144 different matrix elements. However, it is easy to see that only a small number of them is nonzero. In order to find them one can use a graphical technique ilustrated in Table~\ref{tab:hop}.
\begin{table}[t!]
\resizebox{\columnwidth}{!}{
\begin{tabular}{|l|l|l||l|l|l|}
\hline
Initial state & Final State & Label & Initial state & Final State & Label\\
\hline
$\Big| {^{\uparrow}_{\uparrow}} \Big>_{\bf i} \Big| {^{\uparrow}_{\_}} \Big>_{\bf j} $ & \Big| ${^{\uparrow}_{\_}} \Big>_{\bf i} \Big| {^{\uparrow}_{\uparrow}} \Big>_{\bf j} $ & $M_1$ & \Big| ${^{\uparrow}_{\downarrow}} \Big>_{\bf i} \Big| {^{\uparrow}_{\_}} \Big>_{\bf j} $ & \Big| ${^{\uparrow}_{\_}} \Big>_{\bf i} \Big| {^{\uparrow}_{\downarrow}} \Big>_{\bf j} $ & $M_9$\\
\hline
$ \Big| {^{\uparrow}_{\uparrow}} \Big>_{\bf i} \Big| {^{\downarrow}_{\_}} \Big>_{\bf j} $ & $ \Big| {^{\uparrow}_{\_}} \Big>_{\bf i} \Big| {^{\downarrow}_{\uparrow}} \Big>_{\bf j} $ & $M_2$ & $ \Big| {^{\uparrow}_{\downarrow}} \Big>_{\bf i} \Big| {^{\downarrow}_{\_}} \Big>_{\bf j} $ & $ \Big| {^{\uparrow}_{\_}} \Big>_{\bf i} \Big| {^{\downarrow}_{\downarrow}} \Big>_{\bf j} $ & $M_{10}$\\
\hline
$ \Big| {^{\uparrow}_{\uparrow}} \Big>_{\bf i} \Big| {^{\_}_{\uparrow}} \Big>_{\bf j} $ & $ \Big| {^{\_}_{\uparrow}} \Big>_{\bf i} \Big| {^{\uparrow}_{\uparrow}} \Big>_{\bf j} $ & $M_3$ & $ \Big| {^{\uparrow}_{\downarrow}} \Big>_{\bf i} \Big| {^{\_}_{\uparrow}} \Big>_{\bf j} $ & $ \Big| {^{\_}_{\downarrow}} \Big>_{\bf i} \Big| {^{\uparrow}_{\uparrow}} \Big>_{\bf j} $ & $M_{11}$\\
\hline
$ \Big| {^{\uparrow}_{\uparrow}} \Big>_{\bf i} \Big| {^{\_}_{\downarrow}} \Big>_{\bf j} $ & $ \Big| {^{\_}_{\uparrow}} \Big>_{\bf i} \Big| {^{\uparrow}_{\downarrow}} \Big>_{\bf j} $ & $M_4$ & $ \Big| {^{\uparrow}_{\downarrow}} \Big>_{\bf i} \Big| {^{\_}_{\downarrow}} \Big>_{\bf j} $ & $ \Big| {^{\_}_{\downarrow}} \Big>_{\bf i} \Big| {^{\uparrow}_{\downarrow}} \Big>_{\bf j} $ & $M_{12}$\\
\hline
$ \Big| {^{\downarrow}_{\downarrow}} \Big>_{\bf i} \Big| {^{\downarrow}_{\_}} \Big>_{\bf j} $ & $ \Big| {^{\downarrow}_{\_}} \Big>_{\bf i} \Big| {^{\downarrow}_{\downarrow}} \Big>_{\bf j} $ & $M_5$ & $ \Big| {^{\downarrow}_{\uparrow}} \Big>_{\bf i} \Big| {^{\uparrow}_{\_}} \Big>_{\bf j} $ & $ \Big| {^{\downarrow}_{\_}} \Big>_{\bf i} \Big| {^{\uparrow}_{\uparrow}} \Big>_{\bf j} $ & $M_{13}$\\
\hline
$ \Big| {^{\downarrow}_{\downarrow}} \Big>_{\bf i} \Big| {^{\uparrow}_{\_}} \Big>_{\bf j} $ & $ \Big| {^{\downarrow}_{\_}} \Big>_{\bf i} \Big| {^{\uparrow}_{\downarrow}} \Big>_{\bf j} $ & $M_6$ & $ \Big| {^{\downarrow}_{\uparrow}} \Big>_{\bf i} \Big| {^{\downarrow}_{\_}} \Big>_{\bf j} $ & $ \Big| {^{\downarrow}_{\_}} \Big>_{\bf i} \Big| {^{\downarrow}_{\uparrow}} \Big>_{\bf j} $ & $M_{14}$\\
\hline
$ \Big| {^{\downarrow}_{\downarrow}} \Big>_{\bf i} \Big| {^{\_}_{\downarrow}} \Big>_{\bf j} $ & $ \Big| {^{\_}_{\downarrow}} \Big>_{\bf i} \Big| {^{\downarrow}_{\downarrow}} \Big>_{\bf j} $ & $M_7$ & $ \Big| {^{\downarrow}_{\uparrow}} \Big>_{\bf i} \Big| {^{\_}_{\uparrow}} \Big>_{\bf j} $ & $ \Big| {^{\_}_{\uparrow}} \Big>_{\bf i} \Big| {^{\downarrow}_{\uparrow}} \Big>_{\bf j} $ & $M_{15}$\\
\hline
$ \Big| {^{\downarrow}_{\downarrow}} \Big>_{\bf i} \Big| {^{\_}_{\uparrow}} \Big>_{\bf j} $ & $ \Big| {^{\_}_{\downarrow}} \Big>_{\bf i} \Big| {^{\downarrow}_{\uparrow}} \Big>_{\bf j} $ & $M_8$ & $ \Big| {^{\downarrow}_{\uparrow}} \Big>_{\bf i} \Big| {^{\_}_{\downarrow}} \Big>_{\bf j} $ & $ \Big| {^{\_}_{\uparrow}} \Big>_{\bf i} \Big| {^{\downarrow}_{\downarrow}} \Big>_{\bf j} $ & $M_{16}$\\
\hline
\end{tabular}}
\caption{A graphical technique illustrating how to find all nonzero matrix elements for the two-site Hamiltonians (\ref{hamt}-\ref{hamt2}). We have extended the basis and split the $ \frac{1}{\sqrt{2}} ( \Big| {^{\uparrow}_{\downarrow}} \Big> + \Big| {^{\downarrow}_{\uparrow}} \Big> ) $ state into the two `classical' states $\big| ^{\uparrow}_{\downarrow} \big>$ and $\big| ^{\downarrow}_{\uparrow} \big>$. After finding all nonzero matrix elements one needs to project those back onto the triplet state. In this basis the action of the Hamiltonian is simply to move one spin (i.e. one arrow) to the empty orbital on the right. For each initial state there is therefore only one final state. It is then enough to consider the action of the Hamiltonian on all possible initial states, as shown in the table above. Each of these matrix elements has its conjugate, which represents the inverse process.}
\label{tab:hop}
\end{table}
 
The matrix elements listed above and their complex conjugates constitute all nonzero matrix elements. In total there is 32 of them. There are, however, some symmetries that help us write the Hamiltonian in a simpler form.

First, $(M_1,M_2,M_5,M_6,M_9,M_{10},M_{13},M_{14})$ and conjugates come from $(H_t^{xz})_{{\bf i},{\bf i}+{\bf e}_{xz}}$, the rest comes from $(H_t^{yz})_{{\bf i},{\bf i}+{\bf e}_{yz}}$. We only need to consider one of these Hamiltonians as the matrix elements of other one can be obtained in an analogous way. 
From now on we will only consider $(H_t^{xz})_{{\bf i},{\bf i}+{\bf e}_{xz}}$.

Second, let us look at $M_4$ and $M_{11}^*$ or $M_{11}$ and $M_{4}^*$, where a star denotes complex conjugation. They represent exactly the same process but happening in the opposite direction on the lattice. $M_4$ and $M_{11}$ on the other hand represent the inverse processes happening in opposite directions on the lattice. This means we need only calculate one of these matrix elements. The same is true for the pair $(M_8,M_{16})$  and their conjugates. This leaves us with $(M_3,M_4,M_7,M_8,M_{12},M_{15})$.

Third, $(M_3,M_7)$ are the same but have all spins flipped, which is a symmetry of the Hamiltonian. The same is true for $(M_4,M_8)$ and $(M_{12},M_{15})$. Altogether, this leaves us with the classification of the nonzero matrix elements shown in Table~\ref{tab:second}.

\begin{table}[t!]
\resizebox{\columnwidth}{!}{
%\begin{center}
\begin{tabular}{|l|l|l|}
\hline
Matrix elements & Representative\\
\hline
$M_1$, $M_3$, $M_5$, $M_7$ + c.c. & $M_3$\\
\hline
$M_2$, $M_4$, $M_6$, $M_8$, $M_{10}$, $M_{11}$, $M_{13}$, $M_{16}$ + c.c. & $M_4$\\
\hline
$M_9$, $M_{12}$, $M_{14}$, $M_{15}$ + c.c. & $M_{15}$\\
\hline
\end{tabular}
%\end{center}
}
\caption{All nonzero matrix elements of the Hamiltonians (\ref{hamt}-\ref{hamt2}) divided into symmetry classes.}
\label{tab:second}
\end{table}

In order to obtain the second quantised form of (\ref{hamt}-\ref{hamt2}) we need to calculate all matrix elements 
after which we can construct the projected two-site Hamiltonian using
\begin{equation} \label{conc ham}
  (H_t^{xz})_{{\bf i},{\bf i}+{\bf e}_{xz}} = \sum\limits_{k,l} (H_t^{xz})_{{\bf i},{\bf i}+{\bf e}_{xz}}^{k,l} \left| k \right> \left< l \right|,
\end{equation}
where the vectors $\left| k \right>, \left| l \right>$ are written in the second quantised form.
%% KW Please verify that the following equation is correct
$(H_t^{xz})_{{\bf i},{\bf i}+{\bf e}_{xz}}^{k,l} \equiv M_j$ 
are the matrix elements that have been discussed above and need to be explicitly calculated, see below. 

\subsubsection{Calculation of the matrix elements}

The matrix elements $M_j$ can be calculated in a straightforward way. As an example we calculate $M_3$:
\begin{align}
  M_{3} & =\bigg(\left< 0 \right| \tilde{c}_{{\bf i},yz,\uparrow}\tilde{c}_{{\bf j},xz,\uparrow}\tilde{c}_{{\bf j},yz,\uparrow} \bigg) \bigg( -t \: \tilde{c}_{{\bf j},xz,\uparrow}^{\dag}\tilde{c}_{{\bf i},xz,\uparrow} \bigg) \nonumber \\
  & \times \bigg( \tilde{c}_{{\bf j},yz,\uparrow}^{\dag}\tilde{c}_{{\bf i},yz,\uparrow}^{\dag}\tilde{c}_{{\bf i},xz,\uparrow}^{\dag}\left| 0 \right>\bigg)\nonumber \\
& =t \: \left< 0 | 0 \right>.
\end{align}
$M_4$ and $M_{15}$ are calculated in a similar manner. The result is shown in Table~\ref{tab:result}.
\begin{table}[t!]
\resizebox{\columnwidth}{!}{
%\begin{center}
\begin{tabular}{|l|l|l|}
\hline
Matrix elements & Rep. & Value\\
\hline
$M_1$, $M_3$, $M_5$, $M_7$ + c.c. & $M_3$ & $t$\\
\hline
$M_2$, $M_4$, $M_6$, $M_8$, $M_{10}$, $M_{11}$, $M_{13}$, $M_{16}$ + c.c. & $M_4$ & $\frac{t}{\sqrt{2}}$\\
\hline
$M_9$, $M_{12}$, $M_{14}$, $M_{15}$ + c.c.  & $M_{15}$ & $\frac{t}{2}$\\
\hline
\end{tabular}
%\end{center}
}
\caption{The calculated values of matrix elements in each symmetry class.}
\label{tab:result}
\end{table}
Once all the matrix elements are calculated one can easily write down the projected form of Eq.~(\ref{model2}) in the main text in the second quantized form.

\subsubsection{Polaronic mapping}

In order to map our model onto a polaronic one, we need to introduce slave fermions (cf. Ref~\onlinecite{Martinez1991}) using a general mapping
\begin{equation} \label{slavefermions}
  \begin{array}{l}
    \tilde{c}_{{\bf i},\alpha,\uparrow} \rightarrow h_{{\bf i},\alpha}^{\dag}, \quad \quad
    \tilde{c}_{{\bf i},\alpha,\downarrow} \rightarrow \hat{A} \: h_{{\bf i},\alpha}^{\dag} \: S_{\bf i}^+,\\
  \end{array}
\end{equation}
where $\hat{A}$ is an operator to be determined. The spinless hole operators $h_{{\bf i},\alpha}$ obey the Pauli exclusion principle and the standard anticommutation relations. 
The spin $S=1$ operators obey the standard commutation relations. Finally, the spinles hole operators commute with the spin operators, 
which introduces an extra term in the Hamiltonian (see below).

\subsubsection{Finding ${\bf \hat{A}}$}

After transformation (\ref{slavefermions}) the on-site Fock basis consists of the spinless holes with two orbital flavors $xz, yz$ and three eigenvalues of the projection of the spin  
$S=1$ onto the $\hat{z}$ axis ($S_z$). Thus, we label these states by the number of spinless holes on each orbital $n_{xz}, n_{yz} \in \{ 0,1\}$ and the $S_z \in \{-1,0,1\}$ spin quantum number and
write the basis states as $\left\{ \left| n_{xz}, n_{yz}, S_z \right> \right\}$.

To see how the transformation works let us look at the state $\left| {^-_{\downarrow}} \right>$ ($\left| 0 \right> \: \equiv \: \left| 1,1,1 \right>$ defines
the vacuum):
  \begin{align} \label{example state}
      \Big| {^-_{\downarrow}} \Big> &= \sqrt{2} \tilde{c}_{xz , \uparrow} \left( \frac{1}{\sqrt{2}}S^{-} \right) \tilde{c}_{yz , \uparrow}^{\dag} \tilde{c}_{xz , \uparrow}^{\dag} \left| 0 \right> \nonumber\\
      &= \sqrt{2} h_{xz}^{\dag} \left( \frac{1}{\sqrt{2}}S^{-} \right) h_{yz}h_{xz} \left| 1,1,1 \right> = \sqrt{2} \left| 1,0,0 \right>.
  \end{align}
Thus, the state $\Big| {^-_{\downarrow}} \Big>$ maps to the state $\sqrt{2} \left| 1,0,0 \right>$---a state of one hole and one magnon. Similarly, one can determine the other states
\begin{equation} \label{basis map}
  \begin{array}{ll}
    \Big| 0 \Big> \equiv \left| 1,1,1 \right>, \; & \Big| {^{\uparrow}_-} \Big> \equiv \left| 0,1,1 \right>,\\
    \Big| {^{\uparrow}_{\uparrow}} \Big> \equiv \left| 0,0,1 \right>, \; & \Big| {^-_{\uparrow}} \Big> \equiv \left| 1,0,1 \right>,\\
    \Big| {^{\uparrow}_{\downarrow}} \Big> \equiv \left| 0,0,0 \right>, \; & \Big| {^{\downarrow}_-} \Big> \equiv \sqrt{2} \left| 0,1,0 \right>,\\
    \Big| {^{\downarrow}_{\downarrow}} \Big> \equiv \left| 0,0,-1 \right>, \; & \Big| {^-_{\downarrow}} \Big> \equiv \sqrt{2} \left| 1,0,0 \right>.\\
  \end{array}
\end{equation}
We stress that in the above notation the first two quntum numbers are the number of holes on each orbital. [For example, the state $\left| 1, 1, 1 \right>$ is the vacuum (no electrons) and the state $\left| 0, 0, 0 \right>$ is the $S_z = 0$ two electron state.]. The operator $\hat{A}$ is necessary to normalize the states in the new basis. In (\ref{basis map}) we see that two states are not normalized and acquire a factor of $\sqrt{2}$, which is a consequence of the projection onto the $S_z=0$ triplet state. It is easy to check that consequently $\hat{A} = 1$ for $\{ \left|1,0,0\right>, \left|0,1,0\right> \}$ and $\hat{A} = \frac{1}{\sqrt{2}}$ otherwise.

\subsubsection{Restricting the Hilbert space}

Looking at (\ref{basis map}) again, we observe that there are four states that do not map to any states in the old basis, namely
\begin{equation}
  \left\{ \left| 1,0,-1 \right> , \left| 0,1,-1 \right> , \left| 1,1,0 \right> , \left| 1,1,-1 \right> \right\}.
\end{equation}
Evidently, these need to be projected out. One could achieve this using projection operators, but it would complicate the formula for the Hamiltonian. Another approach, presented in Ref.~\onlinecite{Martinez1991} for the $S=1/2$ case, is to include an extra term in the Hamiltonian with a very large coupling constant $\zeta > 0$, in the spirit of the Lagrange multipliers. In our case this term takes the form
\begin{align} \label{constraint}
  \begin{split}
    H_{\zeta} &= \zeta \sum\limits_{\bf i} \Big[ \left( h_{{\bf i} , xz}^{\dag}h_{{\bf i} , xz} h_{{\bf i} , yz}^{\dag}h_{{\bf i} , yz} \left( S_{\bf i}^z \right)^2 \right) +\\
      &+ \left( h_{{\bf i} , xz}^{\dag}h_{{\bf i} , xz} + h_{{\bf i} , yz}^{\dag}h_{{\bf i} , yz} \right)  \left( S_{\bf i}^z - 1 \right) S_{\bf i}^z \Big].\\
    \end{split}
\end{align}
Following the authors of Ref.~\onlinecite{Martinez1991} we will neglect this part of the Hamiltonian. It is clear that this is not without consequence. For simpler models it was shown~\cite{Bieniasz2018} that including such constraints in the diagrammatic expansion of the Dyson equation leads to quantitative differences. We believe that the same situation happens 
for the $S=1$ case studied here. 

\subsubsection{The linear spin wave (LSW) approximation}

To arrive at the formula (5) in the main text we need to find the expressions for the operators $\big| k \big> \big< l \big|$ appearing in Eq.~(\ref{conc ham}). We look for them in the LSW approximation. 

After introducing magnons via the Holstein-Primakoff transformation, the spin quantum number $S_z$ maps onto the number of magnons quantum number $n_{mag}$:
\begin{equation}
    S_z= 1,0,-1  \rightarrow n_{mag}= 0,1,2 \ ,
\end{equation}
respectively, while the fermionic quantum numbers $\{ n_{xz}, n_{yz} \}$ remain the same. 

In this basis, let us examine the projection operator ($P_3$) associated with the matrix element $M_3$ that was discussed above (the other cases are analogous, see below). After the sublattice rotation we obtain
\begin{align}
    P_{3} &= \Big| {^{-}_{\uparrow}} \Big>_{{\bf i}} \Big| {^{\downarrow}_{\downarrow}} \Big>_{{\bf j}} \Big< {^{\uparrow}_{\uparrow}} \Big|_{{\bf i}} \Big< {^{-}_{\downarrow}} \Big|_{{\bf j}} = \nonumber \\
    &= \sqrt{2} \: \left| 1,0,0 \right>_{\bf i} \left| 0,0,2 \right>_{{\bf j}} \left< 0,0,0 \right|_{\bf i} \left< 1,0,1 \right|_{{\bf j}} = \nonumber \\
    &= \sqrt{2} \: \left( h_{{\bf j},yz} h_{{\bf j},xz} h_{{\bf i},yz} a_{{\bf j}}^{\dag}a_{{\bf j}}^{\dag} \left| 1,1,0 \right>_{\bf i} \left| 1,1,0 \right>_{{\bf j}} \right) \nonumber \\
     &\otimes \left( \left< 1,1,0 \right|_{{\bf i}} \left< 1,1,0 \right|_{{\bf j}} a_{{\bf j}} h_{{\bf j},yz}^{\dag} h_{{\bf i},xz}^{\dag} h_{{\bf i},yz}^{\dag} \right). 
\end{align}
First, we notice that the projection onto the double vacuum state, which represents two empty sites, is obsolete. Indeed, if a state survives the action of the spinless fermion creation operators on the right it survives it as one of two states:
\begin{enumerate}
\item A state with four spinless holes, two on each site, in which case the projection is obsolete as this is a unique property of the vacuum,
\item A state with three spinless holes, two on the $i$-th site and one on the $i+1$-st site. In this case the annihilation operators on the left annihilate it, because two of them act on the 
$i+1$-st site.
\end{enumerate}
It is easy to see that the same is true for any of the 16 operators multiplying the matrix elements in Table \ref{tab:hop} and their Hermitean conjugates.

Using this we can write $P_3$ as
\begin{align}
    P_{3} &= \sqrt{2} \: h_{{\bf j},yz} h_{{\bf j},xz} h_{{\bf i},yz} h_{{\bf j},yz}^{\dag} h_{{\bf i},xz}^{\dag} h_{{\bf i},yz}^{\dag} a_{{\bf j}}^{\dag}a_{{\bf j}}^{\dag} a_{{\bf j}} \nonumber \\
  & \approx \sqrt{2} \: h_{{\bf i},xz}^{\dag} h_{{\bf j},xz} a_{{\bf j}}^{\dag}a_{{\bf j}}^{\dag} a_{{\bf j}},
\end{align}
where we have neglected the normal ordered terms with three or more spinless hole operators which go beyond our diagrammatic expansion (see section D). 

We see that $P_3$ is of order three in the bosonic operators. Performing similar calculations for the other 15 operators one can show that they can be divided into three groups:
\begin{enumerate}
\item of order one in bosonic operators,
\item of order three in bosonic operators,
\item of order five in bosonic operators.
\end{enumerate}
In the LSW approximation we only consider the first group of terms. Consequently, only four amongst the 16 operators are non-negligible. These are $ \{ P_2 , P_4, P_{10}, P_{16} \} $. Together with their respective matrix elements the four operators and their Hermitean conjugates give the projected kinetic Hamiltonian in the LSW approximation, see last
two lines of Eq.~(\ref{finalhamiltonian}) in the main text. 

\subsection{Mapping onto a polaronic model: Spectral functions}

As discussed in the main text of the paper we are interested in calculating the following spectra function
\begin{align}
    &{A}_{\alpha}({\bf k},\omega) = -\frac{1}{\pi} {\rm Im}\left\{{G}_{\alpha}({\bf k},\omega) \right\} = \nonumber \\
    &= -\frac{1}{\pi} {\rm Im}  \left\langle 0 \right| \tilde{c}^{\dag}_{{\bf k},\alpha,\sigma} \frac{1}{\omega - \mathcal{H}  + E_0 + i \delta} \tilde{c}_{{\bf k},\alpha\sigma} \left| 0 \right\rangle,
\end{align}
where $\tilde{c}_{{\bf k},\alpha,\sigma} = c_{{\bf k},\alpha,\sigma}(1-c^{\dag}_{{\bf k},\alpha,{\bar \sigma}}c_{{\bf k},\alpha,{\bar \sigma}})$ are the restricted hole annihilation operators. 
It is therefore not a one particle Green's function. 

The relation between the above-defined hole spectral function and the spinless hole spectral function is nontrivial, cf. Appendix of Ref.~\onlinecite{Martinez1991}. 
The latter one, that is natural to the polaronic language, is calculated from the single-particle spinless hole Green's function and reads
\begin{align} \label{spectral ok}
    &A_{\alpha}({\bf k},\omega) = -\frac{1}{\pi} {\rm Im}\left\{ G_{\alpha}({\bf k},\omega) \right\} = \nonumber \\
    &=-\frac{1}{\pi} {\rm Im } \left\langle 0 \right| h_{{\bf k},\alpha} \frac{1}{\omega - H +E_0 + i \delta} h^{\dag}_{{\bf k},\alpha} \left| 0 \right\rangle.
\end{align}
Fortunately, it was shown that for the $S=1/2$ $t$-$J$ model the spinless hole spectral function 
and the hole spectral function almost coincide~\cite{Wang2015}. We assume that the same also
holds also for the $S=1$ $t$-$J$ model investigated here.

\subsection{The self-consistent Born approximation to the Dyson equation} 
\label{sec:scba}

To obtain the Green's function $G_{\alpha}({\bf k},\omega)$ of Eq.~(\ref{spectral ok}),
and thus calculate the spectral function $A_{\alpha}({\bf k},\omega)$,
we use the Dyson equation that reads
\begin{equation}
  G_{\alpha}({\bf k},\omega) = G^{0}_{\alpha}({\bf k},\omega) + G^{0}_{\alpha}({\bf k},\omega) \Sigma_{\alpha}({\bf k},\omega) G_{\alpha}({\bf k},\omega),
\end{equation}
where $\alpha$ is an orbital index. The self energy $\Sigma_{\alpha}({\bf k},\omega)$ is defined as the sum of all non-reducible diagrams starting and ending with the same vertex with an external line representing a spinless hole with momentum ${\bf k}$ and orbital index $\alpha$. 

We calculate the self energy $\Sigma_{\alpha}({\bf k},\omega)$ approximately, using the self-consistent Born approximation (SCBA):
\begin{align} \label{scba diag}
  \begin{split}
    &\Sigma_{\alpha}({\bf k},\omega) \approx \\
    &= \int\limits_{-\infty}^{\infty} d \omega' \sum\limits_{{\bf q}} \; D^{0}(\omega') G_{\alpha}({\bf k} - {\bf q}, \omega - \omega') V_{\alpha} ({\bf k}, {\bf q}) V_{\alpha} ({\bf k}, {\bf q})^*,
  \end{split}
\end{align}
where the vertex is defined as
\begin{align} \label{vertex}
    \begin{split}
    %% KW
%    V_{\alpha} \left( {\bf k}, {\bf q} \right) &= \frac{2 \: t^2}{N} \left| 
%    \left({\gamma}_{{\bf k} \cdot {\bf e}_{\alpha}} v_{\textbf{q}}+{\gamma}_{({\bf k-q}) \cdot {\bf e}_{\alpha}} u_{\textbf{q}} \right) \right|^2\\
        V_{\alpha} \left( {\bf k}, {\bf q} \right) &= \frac{\sqrt{2} \: t}{\sqrt{N}} \
    \left({\gamma}_{{\bf k} \cdot {\bf e}_{\alpha}} v_{\textbf{q}}+{\gamma}_{({\bf k-q}) \cdot {\bf e}_{\alpha}} u_{\textbf{q}} \right), \\
    \end{split}
\end{align}
and the magnon Green's function is
\begin{equation} \label{magnongreen}
    D^{0}(\omega) = \delta(\omega - \Omega_{\bf q}).
\end{equation}

Using Eqs.~(\ref{vertex}) and (\ref{magnongreen}) we obtain the self-consistent equation for the self energy (\ref{scba diag}) in the SCBA approximation
\begin{align} \label{scba}
    &\Sigma_{\alpha}({\bf k},\omega) = \sum\limits_{{\bf q}} \; G_{\alpha}({\bf k} - {\bf q}, \omega - \Omega_{\bf q}) V_{\alpha} ({\bf k}, {\bf q}) V_{\alpha} ({\bf k}, {\bf q})^* \nonumber
    \\
    &= \sum\limits_{{\bf q}} \; \frac{V_{\alpha} ({\bf k}, {\bf q}) V_{\alpha} ({\bf k}, {\bf q})^*}{\omega + J - \Omega_{\bf q} - \Sigma_{\alpha}({\bf k}-{\bf q},\omega - \Omega_{\bf q})}.
\end{align}
Finally, the above equation is solved numerically for the self energy $\Sigma_{\alpha}({\bf k},\omega)$ 
on a finite mesh of ${\bf k}$ and $\omega$ points (see main text of the paper).

%\bibliography{Spinpolruthen}{}
%\bibliographystyle{prsty-etal}

\end{document}